# Semi-Streaming Architecture: A New Design Paradigm for CNN Implementation on FPGAs

Nazariy K. Shaydyuk
Department of Electrical and Computer Engineering
The University of Texas at San Antonio
San Antonio, TX 78249, USA
nazariy.shaydyuk@my.utsa.edu

Eugene B. John
Department of Electrical and Computer Engineering
The University of Texas at San Antonio
San Antonio, TX 78249, USA
eugene.john@utsa.edu

*Abstract*—The recent research advances in deep learning have led to the development of small and powerful Convolutional Neural Network (CNN) architectures. Meanwhile Field Programmable Gate Arrays (FPGAs) has become a popular hardware target choice for their deployment, splitting into two main implementation categories: streaming hardware architectures and single computation engine design approaches. The streaming hardware architectures generally require implementing every layer as a discrete processing unit, and are suitable for smaller software models that could fit in their unfolded versions into resource-constrained targets. On the other hand, single computation engines can be scaled to fit into a device to execute CNN models of different sizes and complexities, however, the achievable performance of one-size-fits-all implementations may vary across CNNs with different workload attributes leading to inefficient utilization of hardware resources. By combing the advantages of both of the above methods, this work proposes a new design paradigm called semi-streaming architecture, where layer-specialized configurable engines are used for network realization. As a proof of concept this paper presents a set of five layer-specialized configurable processing engines for implementing 8-bit quantized MobilenevV2 CNN model. The engines are chained to partially preserve data streaming and tuned individually to efficiently process specific types of layers: normalized addition of residuals, depthwise, pointwise (expansion and projection), and standard 2D convolution layers capable of delivering 5.4GOp/s, 16GOp/s, 27.2GOp/s, 27.2GOp/s and 89.6GOp/s, respectively, with the overall energy efficiency of 5.32GOp/s/W at a 100MHz system clock, requiring total power of 6.2W on a XCZU7EV SoC FPGA.

## I. INTRODUCTION

The focus of the early Convolutional Neural Network (CNN) design and development was mostly on the accuracy of the model. The primary goal was to demonstrate that the image classification performed by CNNs could have a better accuracy than conventional image processing algorithms, or even surpass human-level classification performance [1]. Indeed, by making the models large enough [2], the results obtained were impressive, however, the implementation complexity and usage of such models were becoming an issue. The race to achieve superior model accuracy was introducing larger and deeper network architectures and the computational cost and memory requirements were increasing along with the size and the number of parameters of the models making them impractical for deployment in resource-constrained environments.

The above mentioned limitations led to new ways of CNN modeling which also took into consideration the computational complexity, static and dynamic memory constraints, power consumption and other implementation aspects in order to make CNN models not only accurate but also practical and feasible for deployment. As a result, a new design workflow emerged that focused on model compression [3-5] and the design of smaller and more efficient CNN architectures [6-9] for deploying on embedded systems. FPGAs and SoCs offered an attractive trade-off between the cost, flexibility and performance [10], with their strongest inherent features, viz reconfigurability and fast automated compilation using High Level Synthesis (HLS) tools. The important step in the porting of a CNN model onto the hardware platform is mapping the given software-defined architecture to the corresponding hardware architecture. Such a task is relevant for both automated CNN-to-FPGA converters and manual implementations. Stylianos [11] et al identified two main design-flow tool categories that followed either streaming architecture or single computation engine implementation trends.

The streaming architecture approach allows all the processing to be occurring in parallel at the pixel rate of the camera as if the CNN model is "unrolled". This solution is practical with relatively small CNN models [12], however, medium and large CNN architectures with high number of channels would quickly run out of the limited FPGA resources while trying to meet fully parallel implementation. Alternatively, the hardware footprint of a single computation engine could be adjusted to meet the margin of the available on-chip resources [13-15]. However, creating a universal processing engine to support many CNN architectures and various layer types would eventually approach the performance inefficiencies of a conventional processor. Additionally, such a CNN unit would have to be coordinated by a controller, which is usually the main processor, running other processes and controls the entire system. The extra microcode for managing the functionality of the computation engine could be an unacceptable overhead interfering with the real-time processes run by the processor.

To overcome the shortcomings of the above two extreme opposite methods, this work proposes a new design paradigm called semi-streaming architecture and presents a set of five layer-type specialized computation engines that can be used to implement a representative CNN architecture. Instead of dedicating hardware expensive blocks for every layer as in a true streaming architecture, the semi-streaming data flow implementation would only require several computation engines reused for all the layers of a given network. This approach would permit implementations of larger CNN architectures (like MobileNetV2 [16]) that otherwise would not fit onto an FPGA in their unfolded versions. Furthermore, the layer-specialized engines can be tuned to better fit the processing requirements of the layers and, thus, improve the overall efficiency of the system. Finally, due to the independently streaming dataflow, it could improve the main system processor performance by reducing its interactions with the hardware.

## II. BACKGROUND

This work discusses a semi-streaming architecture design flow by developing a set of computational engines using the representative layers of MobileNetV2 model as an example. The model was selected due to its extreme popularity and efficient architecture, and high suitability for implementation on embedded systems. In fact, the analysis in [17] of representative CNN architectures identifies MobileNetV2 architecture being one of the most efficient architectures in terms of top-1 accuracy density, which is the measure of how efficiently a model uses its parameters. The following sections gives an overview of the characteristics of MobileNetv2 model and briefly discuss hardware aspects inherent to streaming architecture and single computation engine implementation approaches.

### A. Quantized Model

In this paper, benchmark software version of MobileNetV2 – a quantized 8-bit model with input resolution of 224x224 and width multiplier of 1, was adopted for implementation "as is" from the pool of the hosted models by TensorFlow team. They followed the method of linear post-training per-channel quantization of weights and per-layer quantization of activations, as suggested in [18]. This enables the calculation of the quantized result $r$ of the dot product of two quantized vectors using (1), where $a$ and $w$ were the quantized equivalents of real value vectors; $a_0$, $w_0$ and $r_0$ represents the quantized values of real zero, and m1, m2 and m3 are the corresponding real scalars.

$$r = r_0 + \frac{m_1 \cdot m_2}{m_3} \sum (a[i] - a_0) \cdot (w[i] - w_0) \quad (1)$$

The scalars ratio is the only expression that is a non-integer, constant and smaller than 1, and could be converted to an approximated fixed point equivalent by doubling the value until it is in the [0.5, 1) interval which, in turn, could be represented by a truncated integer multiplier *MULT* with an appropriate right *SHIFT*, yielding (2). This step was adopted in our work to match the equivalent integers-only operation used in the TFLite software implementation.

$$r = r_0 + \left[ MULT \sum (a[i] - a_0) \cdot (w[i] - w_0) \right] \gg SHIFT \quad (2)$$

The MobileNetV2 quantized weights are provided as unsigned integers with 8-bits of precision, the bias coefficients as signed 32-bit integers, and the quantization scalars are in double floating-point precision. The bias coefficients do not require 32 bits to be fully represented and may be casted to 16 bits of precision for the standard convolution, *depthwise* and expansion convolution layers and to 18 bits of precision for the projection layers. All double-precision scalars must be converted to their quantized 32-bit unsigned multiplier and 8-bit shift equivalents to comply with (2).

### B. Folding BN Layers into CONV Layers

The activations of each convolutional layer in the model undergo the batch normalization (BN) operation. From the model's organizational perspective, BN has its dedicated layer, since it involves a set of separate parameters which are being learned during the training phase, but are static, just like the weights and the biases, during the inference phase. Since the convolution operation and the batch normalization operation represent linear transformations, both may be combined into a new linear transformation after the training is finished. Fusing the batch normalization layer with its preceding convolutional layer by readjusting the weights and biases of the latter, eliminates the costly calculations at run-time. Consequently, the inference version of the model does not explicitly implement batch normalization layers, as their effects are accounted for during the convolution.

### C. Residual Blocks Architecture

The main body of MobileNetV2 architecture is built by cascading bottleneck residual blocks. The first layer in the block expands the number of channels before performing the *depthwise* convolution in the second layer. The third layer performs the reverse of the first layer by collapsing the high number of dimensions into a tensor to restore the original channel count as that of the input. Therefore, the amount of data entering and exiting the residual block is much smaller than when moving the data internally between the layers. Such bottleneck data flow is extremely favorable for implementation on a general-purpose processor. On the other hand, most of the devices rely on the external memory, which may be in abundance for a CNN application, but limiting the performance due to the bandwidth bottleneck when moving the weights and the intermediate data between the chip and the external memory. Siu et al [19] investigated memory requirement trends across different CNN architectures and recognized activation-bound and weight-bound performance characteristics that varied by two orders of magnitude for weight memory requirements and by three orders of magnitude for activation memory requirements. Furthermore, the peak bandwidth requirements could differ by 40,000x, depending on the CNN architecture. Data movement is also prone for its high power requirements, especially for external DRAMs and may account for over 90% of power consumption [20]. Therefore, the overall efficiency of a system greatly depends on utilization of on-chip memory and the hardware implementation approach.

### D. Streaming Architectures

A streaming CNN architecture approach implies data flow through a chain of sequential hardware blocks. Every layer of the given CNN model is represented by a highly optimized hardware block that is designed to carry out calculations of that particular layer. One of the advantages to this approach is that the blocks are designed with enabled pipelining on their interfaces and allow concurrent execution, resulting in high processing parallelism as illustrated in Fig. 1(a). Additionally, individual tuning of each block ensures optimal performance-resource trade-off and provides highly optimized implementation even at the system level.

This approach also allows control-independent processor-less execution where all the steps and data-flow details are determined at the compile time and may be implemented along with the processing routines. The limiting factor in streaming architectures is the size of the model because the amount of available logic resources quickly depletes with the number of layers. The Haddoc2 tool [12], for example, fully unrolls each layer, including its input and output features and requires all parameters to be stored on chip, while the off-chip transactions only occur for the inputs and the outputs of the model. FpgaConvNet [21], on the other hand, allows run-time device reconfiguration for larger CNN networks, however, for the latency sensitive applications it tends to produce a design that approaches the time-shared, single computation engine paradigm. DeepBurning [22] implements parameterizable blocks and allows inter-layer and intra-layer reuse at run-time.

### E. Single Computation Engines

A single computation engine is a fixed architectural template that is highly configurable and scalable, allowing CNN models of different complexities to target FPGAs of different sizes, as illustrated in Fig. 1(b). Such engine resembles a general-purpose computing unit comprised of alike processing elements performing matrix multiplication operations. As a result, such universal solution implements the most of CNN layers regardless of the size or the function type. Angel-Eye tool [23], for example, is guided by the overall available device resources, while DnnWeaver [24] uses design exploration to select the specific implementation of the CNN. The essential ad-on to the computation engine is the controller, usually a processor, which issues control instructions and performs scheduling of operations. One or more of such units may be run in parallel and independently if enough resources and memory bandwidth are available. The high configurability and control overhead, on the other hand, may yield sub-optimal system performance and may cause inefficient hardware resource utilization.

## III. IMPLEMENTATION METHOD

The benefit of implementing a CNN by following the semi-streaming architecture approach allows streaming data between the computation blocks and enables concurrent pipelined execution of layers as in true streaming architecture. On the other hand, semi-streaming architecture does not require implementation of every CNN layer as individual computation blocks, but instead, employs several layer-specialized computation engines, as opposed to an architecture that follows single computation engine paradigm.

### A. Semi-Streaming Architecture

*Depthwise* and *pointwise* convolution pose different data dependency problems when processing input activation maps. In *depthwise* convolution, each channel may be calculated independently but requires traversing the entire frame in order to produce the result. On the contrary, *pointwise* convolution only needs a single frame pixel to produce the result, but the pixel must comprise the entire depth (all channels) of the input. Chaining similar convolution layers allows maintaining continuous data flow and, thus, permits true streaming architecture. On the other hand, alternating between the *depthwise* and *pointwise* layers, as in the MobilenetV2 model, impedes hardware implementation using true streaming architecture and requires dataflow reordering such that *depthwise* engine gets the entire frame and the pointwise engine gets the entire pixel depth. Consequently, the downstream engine waits for the upstream engine to finish calculating the output while the incomplete result is buffered in the memory. During this time, the dataflow streaming between the engines is halted. This is the basis of semi-streaming architecture.

In this work we take MobileNetV2 as an example target model. To implement the entire pipeline of layers for this model only five computation engines are required. Due to the periodic arrangement of the layers in the soft model, the data can conveniently flow from one engine to another in a circular fashion as depicted in Figure 1. (c). Specifically, each residual block in the model consists of a *pointwise* convolution expansion layer (EXP), a *depthwise* convolution layer (DWC), a *pointwise* convolution projection layer (PRO), and it also includes a layer for adding residual connections (ADD) as needed. As a result, the input data can effectively traverse the layers of the model first by entering the 2D-convoltuion engine (C2D), and then looping through the rest of the engines until completing the entire CNN path as shown in Fig. 1(c). The data is freely streamed across the PRO-ADD-EXP portion of the loop, but must be stopped and reordered when entering and

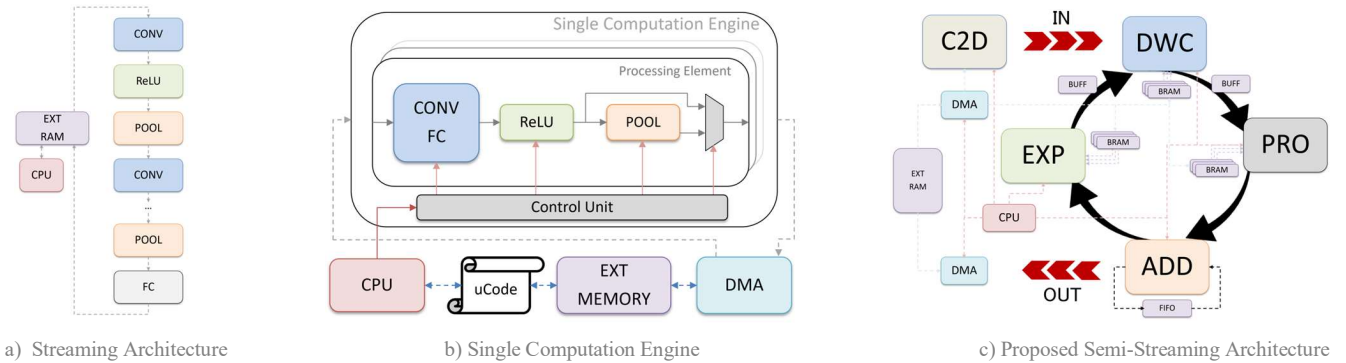

Fig. 1. CNN Implementation Approaches.

a) Streaming Architecture     b) Single Computation Engine     c) Proposed Semi-Streaming Architecture

exiting the DWC engine due to the streaming order incompatibility, which results in a semi-streaming like data flow.

The most common layer types among CNN models are the convolution (CONV), fully connected (FC) and pooling (POOL) which share commonly defined and accepted calculation steps. Therefore, the proposed approach for using layer-specific computation engines with semi-streaming dataflow can be applicable for the most of the CNN models. The design of the computation engines required for implementing MobilenetV2 model is described in the following sections.

*B. C2D Engine: Standard 2D Convolution*

The standard 2D convolution is the most common operation in CNN models and it accounts for over 99% of total operations [25], however, MobileNetV2 model has only two standard convolution layers, whereas only one (entry) layer requires a true 2D convolution for which we adapted the implementation of the C2D specialized engine. The other layer, located right before the average pooling layer, requires 1x1 2D convolution operation, which is identical to *pointwise* convolution.

Spatial 2D convolution requires repeated reads of the same pixel while the convolution window is sliding across the frame in raster fashion. This is not an issue when reading from a memory, however, the true pixel streaming implies volatile dataflow, that is, once the input pixel is read (consumed), it cannot be re-read gain. The work-around for this problem is storing the consumed pixels locally in a line buffer and constantly replacing the old pixels with the new incoming pixels in shift-register like fashion allowing to accept, process and produce new image pixels on every clock cycle.

Each line in the buffer has its dedicated block memory having the width and the depth defined by the number of input channels and the number of columns, respectively. For a sliding window of 3x3, a two-line buffer is sufficient. For example, the first convolution layer in the MobileNetV2 model takes in a 224x224 pixel frame with three 8-bit channels, which would require two 224*3 byte memories. If implemented using discrete size block random access memory (BRAM) resource, such two-line buffer would consume two 512x36bit memory blocks, resulting in about half of the memory resource to be unused. The equations for convolving a single pixel with a filter are represented by (3), (4) and (5):

$$ACC += (a_i - a_0) \cdot (w_i - w_0)$$
$$RES = (ACC \cdot MULT) \gg SHIFT + RES_0$$
$$OUT = RES < MIN \ ? \ MIN : RES > MAX \ ? \ MAX : RES \quad (3\text{–}5)$$

where *ACC* is the sum of the element-wise multiplication of the sliding activations window $a_i$ and the corresponding filter tensor $w_i$ are adjusted to the common zero point reference by $a_0$ and $w_0$, respectively, prior to multiplication; *MULT* and *SHIFT* are the integer-only equivalents for the floating-point multiplier scalar; $RES_0$ adjusts the result with respect to the quantized 0 value, and the *MIN/MAX* clamp the result to enforce [0 255] output activation range. Note that the addition of the bias was omitted for simplification from the current and the future pseudo-code examples, while assuming that the ACC is initialized to the bias value before performing result accumulation.

The equivalent net number of useful convolution operations is 3x3x3 for the convolution and one for the result scaling, totaling 28 and 896 multiply-add (MADD) equivalents per filter per pixel, respectively. The nested loops were completely flattened to allow all per-pixel operations to be executed within one clock cycle with pipelining. Furthermore, the weights and bias arrays were completely partitioned to provide 3x3x3x32+32=896MB/s parameter memory bandwidth.

*C. DWC Engine: Depthwise Convolution*

The *depthwise* convolution is very similar to the standard 2D convolution and also involves sliding a convolution window across two-dimensional space; however, the products are added for each channel separately using the pseudo-code as above, processing the incoming activations in 16-channel batches. The number of channels for every DWC layer is a multiple of 16, so the processing of an n-channel layer takes n/16 passes to complete.

The equivalent net number of useful convolution operations per channel is 3x3 for the convolution and one for the result scaling, totaling 10 and 160 multiply-add (MADD) equivalents per channel and per pixel, respectively. To ensure that the performance is not limited by dada dependency, the weights were stored in 9 separate on-chip memories one for each of the 9 filter pixels. Therefore, the first memory would contain the weights of only the first pixel in filter kernel for all channels and the ninth memory would contain all the weights for the ninth kernel pixel, as shown in Fig. 2(a). Each memory is 128bit wide to allow reading pixel weights for the sixteen channels simultaneously. Similarly, a single bias memory is of 256bit wide which allows to read 16-bit biases for the 16 corresponding channels simultaneously.

The DWC engine could also perform the functionality of the 7-to-1 average pooling layer by taking advantage of already available *depthwise* convolution pipeline and simply skipping the weight multiplication step. Just because the window size for the average pooling is the same size as the input frame (7x7), the function simply accumulates all the frame pixels for each channel and then multiply by the quantized multiplier equivalent to 1/49 to find the average pixel value, similar to (4).

*D. PRO and EXP Engines: Pointwise Convolution*

The pointwise convolution is a special case of the standard 2D convolution which uses a 1x1xChannels kernel size. Therefore, the convolution is performed across the third dimension and requires neither zero-padding nor line buffering. The PRO engine was implemented to perform the projection pointwise convolution operation by following the channels-filters-pixels processing order. That is, all the channels of the first pixel would be convolved with the fist filter, then with the second filter, and so on for all the filters before moving to the next pixel. This processing order would restore the earlier broken pixel streaming dataflow. The pseudo-code is shown in Fig. 3(a), where *APASS* and *FPASS* represent number of n/16 and m/16 passes for the layer with n-channels and m-filters, respectively; *ACT* and *WEI* are the activations and weights batches, and *CLAMP* performs the min/max function of (5). The *HLS PIPELINE* pragma directive is applied to achieve the initiation interval of one for multiplying a 16-channel activation vector by sixteen 16-channel filter vectors simultaneously, which provides a net 256 MADD operations (excluding the calculations servicing loop and memory offset counting). The additional 16 MADD is done for the result scaling at the completion of every 16-filter batch when calculating *RESULT*.

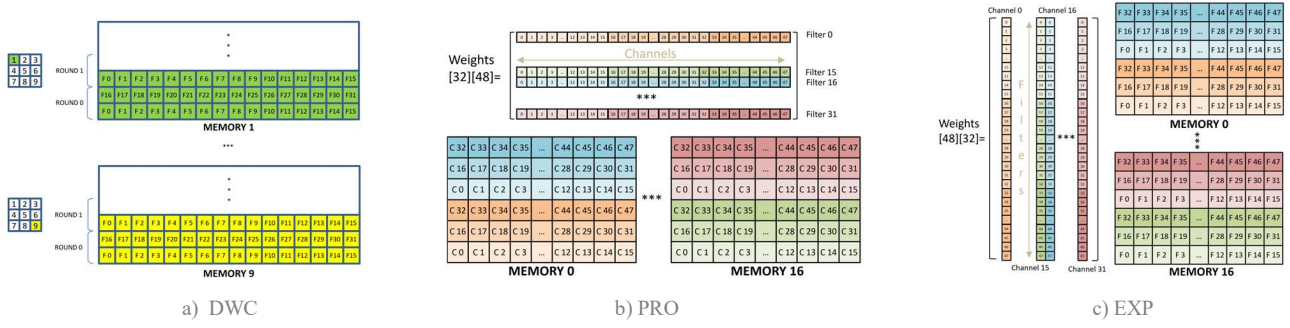

Fig. 2. Weights Memory Organization

The pipelined process with the initiation interval of 1 requires all 16 channels of each of the 16 filters in the batch to be available for calculation simultaneously. To achieve this, 16 separate on-chip memories were dedicated for a batch of 16 filters each containing 16 weight coefficients. Thus, the first memory would contain all the weight coefficients for the first filter of every pass, while the second memory would contain all the weight coefficients of the second filter of every pass and so on, as shown in Fig. 2(b). The bias parameters are stored in a single memory offering 16 coefficients per filter pass. The layers containing number of filters or channels not multiple of sixteen were extended to 32 filters and channels by padding the added locations with quantized zero equivalent values.

The ultimate function of the EXP computation engine is the same as that of the PRO engine, which is to perform *pointwise* convolution. However, the order in which the processing of the input occurs is different than that of the PRO engine and is dictated by the incoming pixel stream. Therefore, as opposed to the PRO engine, which carries the calculations in the channels-filters-pixels order, the EXP engine must follow the filters-channel-pixels processing order. The pseudo-code in Fig. 3(b) shows the loops for *FPASS* and *APASS* with passes swapped.

The pixel stream provides 16 channels of pixel activations which must be convolved with all the corresponding channels of all the filters before consuming the next 16 channels from the stream, since a pixel may be read only once. Therefore, this requires storing temporary convolution results until the last batch of activations is received to finalize the convolution operation. Hence, the filters-channel-pixels processing order requires additional memory resources. The weights organization is different in the EXP memory than in the PRO memory.

The first memory stores the weight coefficients of the first channel of the first filter batch, then the sixteenth channel for the first batch and so on, until all the channels of the first filter batch are covered. Similarly, the second memory holds the second, seventeenth and so on channels of the first batch of filters, as shown in Fig. 2(c). The rest of the batch of filters with all the channels are stored in their corresponding memories. This strategic weight organization allows concurrent read of all required weights for a partial convolution operation. The EXP engine can also perform the 1x1 2D convolution of the layer preceding the pooling layer with no additional hardware resources.

### E. ADD Engine: Normalized Addition of Residuals

The main functionality of the ADD engine represents the residual block shortcuts, which is essentially elementwise addition of the inputs and the outputs. When the residual shortcuts are not present, the block will be configured to pass the stream through to the next processing engine.

The block is accompanied with an AXI-Stream Data FIFO large enough to store all output activations from any projection layer. Therefore, the ADD engine could duplicate the output stream to both the succeeding block and the FIFO to hold the data for retrieval during the next loop when the addition is necessary. Due to the differences in quantization scalars of residual blocks, the input and output activations has to be normalized relative to each other before addition. Therefore, the actual addition of the activation shortcuts is performed after each was scaled as shown in Fig. 4, where *MULT1-3* and *SHIFT1-3* are the quantized integer multipliers with their corresponding right-shifts pre-calculated according to the TensorFlowLite *prepare()* function.

```
for fpass<FPASS
  for apass<APASS
    #pragma HLS PIPELINE II=1
    for f<16
      for a<16
        ACC[ f] += (ACT[apass*16 + a] - az) * (WEI[fpass*16 +f] [apass*16 + a] - wz)
      end
    end
    if apass==APASS-1
      RESULT = (ACC[:] * MULT) >> + oz
      OUT[:] = CLAMP(RESULT[:])
    end
  end
end
```
a) Projection

```
for apass<APASS
  for fpass<FPASS
    #pragma HLS PIPELINE II=1
    for f<16
      for a<16
        ACC[fpass][ f] += (ACT[apass*16 + a] - az) * (WEI[fpass*16 +f] [apass*16 + a] - wz)
        if apass==APASS-1
          RESULT[f] = (ACC[fpass][f] * MULT) >> + oz
          OUT[fpass][f] = CLAMP(RESULT[f])
        end
      end
    end
  end
end
```
b) Expansion

Fig. 3. Pseudo-codes for Pointwise Convolution

## IV. RESULTS AND DISCUSSION

The computation engines were designed, implemented and tested on the Xilinx ZCU104 platform with 100MHz system clock. The Vivado HLS synthesis resource utilization and computational performance for each block is summarized in Table 1.

The C2D computation engine was designed to complete processing of the entry layer by traversing an input frame only once, which allowed maintaining the pixel-rate streaming data flow and it demonstrated the true streaming architecture implementation. This also shows the amount of FPGA resources necessary for implementing a processing block for one relatively small layer (3 input and 32 out channels). If blocks, similar to this, were separately implemented for every individual layer of the network, as required for a true streaming architecture design, the FPGA resources would be quickly depleted. Since the C2D engine is outputting 32 channel pixel stream, while the rest of the blocks supported 16 channels, the output of the engine is split in to two 16 channel streams, with the first 16 channels to be streamed directly to the next processing engine, while the second stream could be routed to the data mover and saved in the activations buffer for later processing.

The design of the DWC engine favors layers with less channels over smaller frame sizes. This is due to the resource consumption trade-off. For example, two *depthwise* convolution layers with the same number of channels, but different frame sizes like 224x224 and 448x448 would require the same amount of DSPs and memory resources because the DSP utilization is dictated by the number of channels to be processed in parallel, while the line-buffer memory utilization is largely dictated by the frame size. Given the discrete sizes of memory building blocks (minimum 512 words), the same memory depth would be allocated for both layers, consuming the same number of BRAM units. This condition would not be necessarily true for the ASIC designs, since the memory sizes could be individually tailored to the required dimensions.

The PRO and EXP engines were designed to perform identical functionality – *pointwise* convolution, however, were implemented to follow different data processing orders. The design of the EXP engine was constrained by the streaming data-flow requirement on its input. As was explained earlier, every consumed element at the input would have to be completely processed by the block before consuming the next element. Consequently, the EXP engine required 15 BRAM18K blocks of additional memory to store the partial results of the filter batch calculations, as is reflected in the resource utilization report. The design of the PRO engine, on the other hand, is not restricted to the specific data processing

```
for r<ROWS
  for c<COLS
    for a<16
      A1 =   (MULT1 * ((IN1[a] – a1z) << 20 ) ) >> SHIFT1
      A2 =   (MULT2 * ((IN2[a] – a2z) << 20 ) ) >> SHIFT2
      RESULT = ((A1+ A2) * MULT3) >>  SHIFT3 + oz
      OUT[a] = CLAMP(RESULT)
    end
  end
end
```

Fig. 4. Normalized Residual Addition

TABLE I. RESOURCE UTILIZATION REPORT

|  | C2D | DWC | PRO | EXP | ADD |
|---|---|---|---|---|---|
| clk period, ns | 3.731 | 4.493 | 3.595 | 3.418 | 3.510 |
| LUTs | 11318 | 7712 | 8483 | 8231 | 9952 |
| FFs | 3521 | 4667 | 7438 | 5627 | 3702 |
| BRAM18k | 2 | 8 | 0 | 15 | 0 |
| DSPs | 608 | 186 | 331 | 324 | 178 |
| GOp/s @100MHz | 89.6 | 16.0 | 27.2 | 27.2 | 5.4 |
| GOp/s/DSP | 0.1474 | 0.8602 | 0.8218 | 0.8395 | 0.3034 |

order and does not consume additional memory resources but requires 7 more DSP blocks than the EXP engine implementation. Therefore, the design of the PRO engine could be optionally selected between the two implementations based on the shortage of a specific FPGA resource or other design factors.

The computation performance of the processing engines also depends on the memory bandwidth of parameter storage. Having the complete set of weights and biases simultaneously available to carry a processing operation in parallel on a given number of channels is an essential requirement for efficient use of the DSP resources. Therefore, the amount of memory for each computation engine is allocated to exactly match the memory bandwidth requirements. The BRAM (BR), URAM (U) memory resource utilization for the weights (W), bias (B) and activations (A) of each computation engine is given in Table II. As explained earlier, the weight and bias parameters for the C2D engine are stored in discrete registers and, thus, do not require memory allocation. It is also worth noting that casting the bias coefficients from 32bit to 16bit precision (18-bit for PRO layers) helps in reducing the amount of the bias memory by one half.

The design for these five computation engines targeted for MobileNetV2 shows high total utilization of the on-chip resources: 94.91% of DSPs, 85.74% of BRAM and 83.33% of URAM units. This also implies high place and route congestion, which Vivado estimated to be level 6 (64x64), likely due to the high net count required for interfacing the Simple Dual Port (SDP) memory instances. Given enough on-chip memory for parameter storage would not require parameter reloading from the external memory, therefore, allowing implementation of read-only memory (ROM). Consequently, the number of nets to route would be cut in half and would likely decrease the congestion level allowing higher processing frequencies. Moreover, the ROMs would be all loaded during bit-stream device configuration which would eliminate the need for accessing external RAM at run-time, freeing out the bandwidth.

All the five computation engines with their accompanying memories, Zynq Processing System, and DMAs were instantiated in the common design and the power consumption was estimated to be 6.197W. Almost 50% of the total power was consumed by the processing system, therefore, the favorable action would be to remove the PS from the design, if it is not being used by the rest of the integrated system, and instead, implement a state machine in the Programmable Logic (PL) to configure computation engines and control the flow of CNN inference process in a processor-less fashion.

TABLE II. MEMORY UTILIZATION AND BANDWIDTH

|  | C2D | DWC | | PRO | | EXP | | ADD | BUFF |
|---|---|---|---|---|---|---|---|---|---|
| Purpose | W,B | W | B | W | B | W | B | A | A |
| Type | - | BR | BR | BR | BR | BR | BR | U | U |
| Bitwidth | - | 128 | 256 | 128 | 288 | 128 | 256 | 128 | 128 |
| Depth | - | 512 | 512 | 1536 | 512 | 2048 | 1024 | 8192 | 77824 |
| #memories | - | 9 | 1 | 16 | 1 | 16 | 1 | 1 | 2 |
| BRAM36k/mem | - | 2 | 4 | 6 | 4 | 7.5 | 7.5 | 4 | 38 |
| BRAM36k total | - | 22 | | 100 | | 127.5 | | 4 | 76 |
| Gb/s, 100MHz | 6928 | 140.8 | | 233.6 | | 230.4 | | 12.8 | 12.8 |

Table III summarizes the computation performance of hardware blocks developed in this work with other automatically created designs and manual implementations. The fairest performance comparison would be when different designs are implemented using the same type of layer for a specific CNN-FPGA pair. However, the available reports tend to use different hardware platforms, CNN models, clocking conditions, etc. Therefore, to provide a more meaningful performance comparison, only the standard 2D convolution block implementations were considered across all designs, while the overall performance [given in brackets] was taken as an average between all the layers.

The first three implementations in the table were produced by [11] when evaluating CNN to FPGA mapping tools: FpgaConvNet, DeepBurning and DnnWeaver, where the first two followed the streaming architecture paradigm, and the last one specialized in the design of single computing engine. Considering the FPGA devices of different number of available DSP slices (220 vs. 1728) and clocking frequencies (125MHz vs. 100MHz) the comparison could be done by normalizing the conditions. For example, if the FpgaConvNet design reduced the system clock to 100MHz and utilized 608 DSPs as in our design, its normalized performance would be approximately 84.67GOp/s, assuming linear relationship of computation performance with clock frequency and resource utilization. Similar comparison using the aforementioned normalization approach for the other two implementations produced by DeepBurning and DnnWeiver mapping tools would yield computation performances of 51.2GOp/s and 37.14GOp/s, respectively.

The manual design implementations use different CNN models, FPGA platforms, clocking frequencies and arithmetic precision. The implementation from [13] uses 32-bit floating point precision, and assuming it requires 3 DSP48 slices per single multiply-accumulate, we may lower the number of DSPs used for MADD operation from 2240 to 746, yielding the normalized performance of 50.22GOp/s. A similar adjustment can be done for the work from [14], which uses 16-bit fixed point precision, assuming 1 DSP per MADD results in the normalized performance of 97.6GOp/s. Finally, the performance after normalizing the conditions of the [15] design implementation yields 84.56GOp/s. Note that considering a low-power embedded system, the design in our work offers the lowest total power consumption. It is also important to emphasize that the precise performance comparison would be possible only if the designs were under the same conditions and most importantly, implemented the same CNN model layers on the same FPGA.

Fig. 5 shows per-round time line in system clock counts which can be converted to time scale based on the clock period. Note that round 0 also includes the C2D engine processing latency. Up to round 10, the time to load the PRO and EXP weights is less than the processing time of the DWC engine, thus, the system is computation limited. As soon as the DWC finishes processing, the PRO-ADD-EXP triplet can start processing immediately. On the other hand, the parameter loading time for the rounds 13 and above, greatly exceeds the DWC engine processing time, and the second processing stage must remain idle before the weights have been loaded. Hence, the system becomes memory bandwidth limited. The computation load is well-balanced for the engine triplet of the second stage: the processing load for the PRO and the EXP engines is almost always equal, which allows the work in parallel; demonstrating the concurrency feature of streaming hardware architecture, therefore, only the EXP processing bar is plotted. The theoretical latency for looping through all MobileNetV2 layers (17 rounds) was estimated to be 10.6ms (~94f/s), which assumes that the bandwidth of the external RAM data bus to be fully available for parameter transfer. The actual latency, however, would depend on the design of the entire embedded system.

## V. CONCLUSION

A semi-streaming architecture with multiple layer-specialized engines was proposed in this work as an alternative approach to the two opposite CNN hardware architecture paradigms. The proposed design enables the implementation of larger CNN models, such as MobileNetV2, with more overall efficient performance. A set of five layer-specific computation

TABLE III. PERFORMANCE AND EFFICIENCY COMPARISON

|  | CNN-To-FPGA Mapping Tools | | | Custom Architectures | | | This Work |
|---|---|---|---|---|---|---|---|
| Tool | FpgaConvNet | DeepBurning | DnnWeaver | Vivado | Vivado | Xilinx SDx | Vivado HLS |
| Year [work] | 2018 [11] | 2018 [11] | 2018 [11] | 2015 [13] | 2016 [14] | 2019 [15] | - |
| CNN Model | AlexNet | AlexNet | AlexNet | Custom | VGG16 | AlexNet | MobileNetV2 |
| FPGA Platform | XC7Z020 | XC7Z020 | XC7Z020 | VX485T | XC7Z045 | XCZU7EV | XCZU7EV |
| Frequency, MHz | 125 | 100 | 150 | 100 | 150 | 300 | 100 |
| Used DSPs | 220 | 220 | 220 | 2240 | 780 | 696 | 608 [1627] |
| Precision, bits | fixed16 | fixed16 | fixed16 | float32 | fixed16 | int8 | uint8 |
| Performance, GOp/s | 38.30 | 18.53 [15.3] | 20.16 [20.51] | 61.62 | 187.8 [137] | 290.40 [14.11] | 89.60 [33.08] |
| Normalized, GOp/s | 84.7 | 51.2 [42.28] | 37.14 [37.78] | 50.22 | **97.6 [71.2]** | 84.56 [4.11] | **89.60 [33.08]** |
| Perform. Density, GOp/s/DSP | 0.171 | 0.084 [0.07] | 0.0916 [0.093] | - | - | - | 0.147 [0.020] |
| Power, W | - | - | - | 18.61 | 9.63 | 17.67 | **6.2** |
| Efficiency, GOP/s/W | - | - | - | 3.31 | 19.50 [14.2] | 16.44 [0.80] | 14.45 [5.34] |

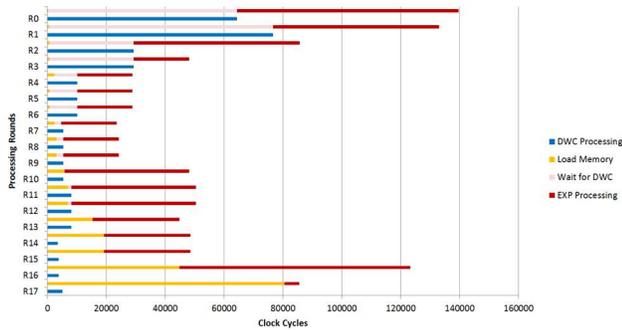
Fig. 5. Per-Round Computation Latency

engines for *depthwise*, *pointwise* (expansion and projection), addition/pooling and standard 2D convolution layers were designed using MobileNetV2 as the target network. The designed engines are capable of delivering 16GOp/s, 27.2GOp/s, 27.2GOp/s, 5.4GOp/s and 89.6GOp/s, respectively, with the overall energy efficiency of 5.32GOp/s/W at a 100MHz system clock, requiring the total power of 6.2W on a XCZU7EV SoC FPGA. The estimated latency for classifying a single image was found to be 10.6ms, that is, over 94 frames per second.

## VI. Future Works

Further research could be done to expand the variety set of layer-specific computation engines for a larger pool of CNN architectures. The implementations from the literature review and this work show trends in maximizing the resource utilization of a target device with the aim of maximizing the performance. Finally, the design space exploration could be performed to find the most efficient implementation.